\documentclass{tcibook}
\usepackage{fancyhea}
\usepackage{work}
\usepackage{bm}       
\usepackage{graphicx}
\usepackage{multirow}

\newcommand{\nc}{\newcommand}  

\def\beq{\begin{equation}}
\def\eeq#1{\label{#1}\end{equation}}
\def\eeqn{\end{equation}}

\newenvironment{Eqnarray}%
   {\arraycolsep 0.14em\begin{eqnarray}}{\end{eqnarray}}
\def\beqa{\begin{Eqnarray}}
\def\eeqa#1{\label{#1}\end{Eqnarray}}
\def\eeqan{\end{Eqnarray}}

\nc{\ra}{\rightarrow}  
\nc{\slsh}{\slash\hspace*{-0.22cm}}
\def\Re{{\cal R \mskip-4mu \lower.1ex \hbox{\it e}\,}}
\def\Im{{\cal I \mskip-5mu \lower.1ex \hbox{\it m}\,}}

\nc{\vev}[1]{ \left\langle {#1} \right\rangle }
\nc{\bra}[1]{ \langle {#1} | }
\nc{\ket}[1]{ | {#1} \rangle }
\nc{\fb}{\,{\rm fb}^{-1}}
\nc{\ev}{{\rm eV}}
\nc{\kev}{{\rm keV}}
\nc{\Mev}{{\rm MeV}}
\nc{\gev}{{\rm GeV}}
\nc{\tev}{{\rm TeV}}
\nc{\mev}{{\rm MeV}}

\def\del{\partial}
\def\Dslash{\not{\hbox{\kern-4pt $D$}}}
\def\dslash{\not{\hbox{\kern-2pt $\del$}}}
\def\pslash{\not{\hbox{\kern-2pt $p$}}}
\def\ETmiss{ \not{\hbox{\kern-4pt $E$}}_T }

\def\msb{{\bar{\ssstyle M \kern -1pt S}}}

\begin{document}

\def\bibname{References}
\bibliographystyle{plain}

\raggedbottom

\pagenumbering{roman}

\parindent=0pt
\parskip=8pt
\setlength{\evensidemargin}{0pt}
\setlength{\oddsidemargin}{0pt}
\setlength{\marginparsep}{0.0in}
\setlength{\marginparwidth}{0.0in}
\marginparpush=0pt


\pagenumbering{arabic}

\renewcommand{\chapname}{chap:intro_}
\renewcommand{\chapterdir}{.}
\renewcommand{\arraystretch}{1.25}
\addtolength{\arraycolsep}{-3pt}

\thispagestyle{empty}
\begin{centering}
\vfill

{\Huge\bf Planning the Future of U.S. Particle Physics}

{\Large \bf Report of the 2013 Community Summer Study}

\vfill

{\Huge \bf Chapter 10: Communication, Education, and Outreach}

\vspace*{2.0cm}
{\Large \bf Conveners: M. Bardeen and D. Cronin-Hennessy}
\pagenumbering{roman}

\vfill

{\large  Study Conveners: M. Bardeen, W. Barletta, L.~A.~T.~Bauerdick, R. Brock,
D.~Cronin-Hennessy, M.~Demarteau, M.~Dine, J.~L. Feng, M. Gilchriese,
S. Gottlieb, J.~L.~Hewett, R. Lipton, H.~Nicholson, M.~E. Peskin,
S. Ritz, I.~Shipsey, H. Weerts}\\
\vspace{1cm}

{\large Division of Particles and Fields Officers in 2013:
J.~L. Rosner (chair), 
I. Shipsey (chair-elect), 
N. Hadley (vice-chair),
P. Ramond (past chair)}\\
\vspace{1cm}

{\large Editorial Committee:
R.~H. Bernstein,
N. Graf,
P. McBride,
M.~E. Peskin,
J.~L. Rosner,
N.~Varelas,
K. Yurkewicz}

\vfill

\end{centering}

\newpage
\mbox{\null}

\vspace{3.0cm}

{\Large \bf Authors of Chapter 10:}

\vspace{2.0cm}
 {\bf  M. Bardeen,  D. Cronin-Hennessy}, 
R. M. Barnett, 
P. Bhat, 
K. Cecire, 
K. Cranmer, 
T. Jordan,
I.~Karliner, 
J. Lykken, 
P.~Norris, 
H.~White, 
K.~Yurkewicz

 \tableofcontents

\newpage

\pagenumbering{arabic}

%


\setcounter{chapter}{9}

\chapter{Communication, Education, and Outreach}
\label{chap:ceo}

\begin{center}\begin{boldmath}



\begin{center}

\begin{large} {\bf Conveners:  M. Bardeen,  D. Cronin-Hennessy} \end{large}

R. M. Barnett, 
P. Bhat, 
K. Cecire, 
K. Cranmer, 
T. Jordan,
I. Karliner, 
J. Lykken, 
P.~Norris, 
H.~White, 
K.~Yurkewicz

\end{center}



\end{boldmath}\end{center}

\section{Introduction}

The U.S. particle physics community recognizes the critical importance of consistent and coherent communication, education, and outreach (CE\&O) activities. These efforts foster nationwide support for the field, develop the next generation of researchers, and help develop a scientifically literate citizenry. 

Federally supported research in particle physics and related fields has led to an impressive list of Nobel Prize-winning discoveries: the first detailed study of the cosmic microwave background; the discovery of neutrino masses and mixing (and earlier work on solar neutrinos many years earlier); the discovery of the accelerating expansion of the universe; understanding the strong force; and the discovery that the Cabbibo-Kobayashi-Maskawa (CKM) matrix explains experiments that have observed an asymmetry in the behavior of particles and anti-particles. Most recently, the 2013 Nobel Prize was awarded for the discovery of the Higgs boson, an achievement made possible by scientific talent, technology and leadership from the United States. 

The American public is fascinated by these discoveries, and by the full breadth of current and future particle physics projects. The saga of the Large Hadron Collider (LHC) and Higgs boson discovery reached audience levels unprecedented for a particle physics event. Public lectures and other events on particle physics topics draw crowds. Milestones, discoveries, and even proposals for projects in particle physics routinely make headlines. 

Many individuals, groups and institutions in the U.S. particle physics community reach out to members of the public, decision makers, teachers, and students through a wide variety of activities. These activities may be effective in achieving their individual goals. However, there is room for improvement in the nationwide coordination of these activities, in the mobilization of the entire U.S. community to take part in the activities, and in efforts to use varied activities to convey consistent and compelling messages to stakeholders.

Translating this public excitement into greater support for the field requires a greater fraction of our community to engage in CE\&O activities, and for the quality, coordination and number of existing activities to be increased. Our community must lower barriers to engagement in such activities, moving towards treating CE\&O activities as another necessary component of research along with publishing and presenting scientific results.  We must also increase our efforts to convey consistent, coherent, and compelling messages about the importance of particle physics research and its value to society. National coordination, training, and additional supporting resources will greatly help support an increased commitment to effective CE\&O. 

This Snowmass process was the first in which one working group was devoted to CE\&O. Our task was threefold: to write a paper that summarizes current activities and provides recommendations for future activities; to carry out CE\&O activities at the Minnesota meeting; and to educate and train physicists at Snowmass about CE\&O. The Snowmass CE\&O group was composed of five working groups organized by the main target audiences for U.S. CE\&O activities: the general public; policy makers and opinion leaders; the science community; teachers in grades 5-16; and students in grades 5-16. As part of the Snowmass Process, the CE\&O group held two pre-meetings associated with the March and April meetings of the American Physical Society, carried out formal and informal discussions with colleagues, surveyed the particle physics community regarding participation in and attitudes toward CE\&O~\cite{PMOLsubgroup}, and established a new website to update the 2005 particle physics education and outreach activities database. During {\it CSS 2013 Snowmass on the Mississippi}, we held lunchtime discussions with over 190 attendees and organized two panels\textemdash {\it Selling Long-Term Science in Washington} and {\it Fostering Interconnections and Common Cause with Scientists from Other Fields}. The particle physics magazine {\it symmetry} invited physicists to explain why particle physics matters in one-minute video talks and compiled a photo gallery of physicists holding whiteboards showing the big question in particle physics that drives their research. 

Several common themes emerged from the CE\&O study as necessary to take particle physics CE\&O in the United States to a new level: 
\begin{itemize}
   \item Making a {\it coherent} case for particle physics\textemdash the compelling questions we address, the facilities we need for our research, and the value of particle physics to society.
   \item Recognizing\textemdash formally and informally\textemdash physicists, postdocs, and students who devote time to CE\&O efforts. 
   \item Developing and increasing access to resources, training activities, and opportunities that engage physicists with policy makers, opinion leaders, the general public, educators and students.
   \item Creating a national team dedicated to developing and providing communication and education strategy and resources, supporting and enhancing existing efforts, and to mobilizing a greater fraction of the U.S. community to participate in CE\&O activities.
\end{itemize}

This summary report begins with a set of overarching goals for U.S. particle physics CE\&O identified as a result of the Snowmass study, along with a set of implementation recommendations that support the goals for all target audiences. The balance of the report is then divided into four audience groups: policy makers and opinion leaders; the general public; the science community; and teachers and students in grades 5-16. Each of the four sections provides an overview of existing CE\&O efforts targeted at that audience group, strategies to achieve the overarching CE\&O goals with that audience, and recommendations for implementation of the identified strategies. More detailed information for all audience groups can be found in the full subgroup reports.

\section{Goals}

The CE\&O group developed its goals, strategies and recommendations with input from particle physicists and education and outreach professionals obtained prior to and during the Minneapolis meeting. The recommendations support a proactive, coordinated CE\&O effort from the entire U.S. particle physics community, reinforcing the view expressed in the 2013 European Strategy Report: ``Outreach and communication in particle physics should receive adequate funding and be recognized as a central component of the scientific activity"~\cite{EUStrategy2013}.  The following are the three overarching goals identified for all U.S. particle physics CE\&O activities:

\begin{enumerate}
 	\item Ensure that the U.S. particle physics community has the resources necessary to conduct research and maintain a world leadership role.
	\item Ensure that the U.S. public appreciates the value and excitement of particle physics.
	\item Ensure that a talented and diverse group of students enter particle physics and other STEM careers, including science teaching.
\end{enumerate}
 
The CE\&O group identified the following four actions that could be taken over the next five years to achieve these goals for all target audiences:

\begin{enumerate}
 	\item Augment existing CE\&O efforts with additional personnel and resources dedicated to nationwide coordination, training and support.
	\item Develop a comprehensive central communication, education, and outreach resource for physicists, with initial content available before the end of the 2013/2014 P5 process.
	\item Provide communication training to the U.S. particle physics community.
	\item Work with the APS Division of Particles and Fields (DPF) and the High Energy Physics Advisory Panel (HEPAP) to develop a sustainable process for collecting statistics on workforce development and technology transfer and with APS to investigate a U.S. economic impact study for physics research that includes particle physics.
\end{enumerate}

The following four sections discuss additional strategies to achieve the overarching goals for specific target audiences.

\section{Building support among policy makers and opinion leaders}

The U.S. particle physics community has recognized as part of the Snowmass process that it must embark on a coordinated effort that mobilizes a greater fraction of scientists and students to translate the public excitement about and interest in particle physics research into greater support among the policy makers that make decisions about research funding and the opinion leaders whose views they trust to guide them in their decisions.

The U.S. particle physics community has engaged for decades in a variety of efforts to educate and inform the public, including policy makers and opinion leaders, about the excitement and importance of particle physics research and its benefits to society. These public outreach efforts have met with success, as demonstrated by the public attention to events in particle physics over the last six years, including the saga of the Large Hadron Collider, experimental hints of faster-than-light neutrinos, and the transit of a huge electromagnet from New York to Chicago. Public lectures and other events on particle physics topics draw crowds that other fields of science envy. Discoveries, milestones and even project proposals in particle physics routinely make headlines, while successes in other fields of science fail to attract media attention.

Yet while particle physics research fascinates and excites the American public, their continuing support for the funds required to build new facilities is not guaranteed. In the current climate of fiscal austerity, government-funded programs must make compelling cases for the societal benefit of their work. The bar is set high for scientific fields such as particle physics, which receives more than \$750 million each year in the budget of the DOE Office of High Energy Physics and has proposed to build new projects with total costs of more than \$1 billion.

A CE\&O survey of more than 600 respondents conducted in the first half of 2013 revealed that approximately 60\% of particle physicists are engaged in outreach activities to the general public; 50\% are engaged in activities that reach K-12 teachers or students; 35\% in activities that reach scientists in other fields; and only 30\% in activities that reach policy makers and opinion leaders~\cite{PMOLsubgroup}. Building greater support among policy makers will require a larger fraction of the particle physics community to engage in these activities, and for these activities to be coordinated, consistent, and effective.

The summary for this subgroup of the CE\&O frontier contains strategic recommendations for actions the U.S. particle physics community can take over the next few years to improve communication and outreach to policy makers and opinion leaders. Policy makers include elected and non-elected officials in federal, state and local government who influence particle physics funding. Opinion leaders are defined as notable figures whose views on scientific research and science funding influence policy makers and the public.

The full subgroup report~\cite{PMOLsubgroup} includes a definition of the two audiences, a more comprehensive overview of existing activities targeted to those audiences, and additional details about each of the recommended activities.

\subsection{Overview of existing activities}
The U.S. particle physics community engages in a number of efforts to build support for research among policy makers and opinion leaders, including: 

\begin{enumerate}
 	\item Annual visits to Washington, D.C. by facility user groups
	\item Email and letter-writing campaigns at key points in the budget cycle
	\item Participation by particle physicists in Washington, D.C. events organized by AAAS, APS, the National User Facilities Organization and other organizations
        \item Direct advocacy with legislators by scientific and industrial societies, national laboratories and individual scientists
        \item Efforts to place stories of particle physics and particle physicists in influential media outlets.
\end{enumerate}

\subsection{Communication and outreach to policy makers and opinion leaders}

We identified three strategies that will help achieve the first two overarching CE\&O goals. These strategies are high-priority efforts that should be undertaken by the particle physics community over the next five years.
\begin{itemize}
 	\item Empower and enable members of the particle physics community to communicate and advocate coherently, consistently and effectively on behalf of their science.
	\item Develop an enduring process to track, update, and disseminate statistics on the impact of particle physics on society.
	\item Put informed third-party advocates to work raising the profile of and informing key stakeholders about the importance of particle physics, physics, and discovery science to the United States.
\end{itemize}

These strategies are targeted to the policy maker and opinion leader audiences, but also support the goals of outreach to the general public and scientists in other fields. They are in turn supported by strategies identified by other CE\&O subgroups, including: 
\begin{itemize}
   \item Develop consensus in our field that we need to prioritize, buy into the mechanism of prioritization, and support the resulting plan. 
   \item Make the public aware of direct and indirect applications of research, both historical and potential.
   \item Communicate the role and stories of U.S. physicists in particle physics, particularly in major discoveries and in the context of our international collaborations.  
   \item Foster more dialog and understanding among subfields of physics and with other sciences. Identify areas of common cause and unite in support of them.  
\end{itemize}

\subsection{Implementing the strategies}

The U.S. particle physics community has been engaged for decades in a variety of efforts to inform the public and policy makers about particle physics research, and to encourage support for that research. The activities recommended below do not replace existing efforts. Instead, they augment and enhance ongoing efforts by providing nationwide coordination and support, and by developing needed resources to make a compelling case for support of particle physics research.

\textbf{Recommendation 1: Augment existing efforts with additional personnel and resources dedicated to nationwide coordination, training and support.}

The U.S. particle physics community, as part of the 2013 Snowmass process, has strongly indicated that it wishes to greatly enhance its CE\&O efforts. Taking things ``to the next level" requires a nationwide effort that mobilizes a greater fraction of the community and is supported by dedicated personnel and resources. This effort will need to be a partnership between scientists and professionals in the areas of communication, government relations, and education. It will need to encompass existing efforts carried out at laboratories, universities and by experimental collaborations.

The greatest initial need is a team of a few people to coordinate and support existing activities and spearhead and organize new nationwide initiatives. Such a team would be most effective if coupled closely with existing nationwide efforts, such as the particle physics magazine {\it Symmetry}, QuarkNet, the outreach activities of the Contemporary Physics Education Project, and the advocacy efforts of users' groups. Funding would also be required to establish and maintain technical infrastructure for the national effort, such as websites and databases.

\textbf{Recommendation 2: Develop a central communication, outreach, and education resource for physicists, with initial content available before the end of the 2013/2014 P5 process.}

The first activity for the team identified in Recommendation 1 would be the creation of a website to act as physicists' central clearinghouse for resources related to CE\&O. This website would be designed to fulfill the needs of physicists who engage in CE\&O activities, not the needs of the general public who wish to learn about particle physics.  It would provide physicists with the tools, techniques, information, and resources they need to engage effectively with policy makers, opinion leaders, the general public, teachers, and students. The content would include:
\begin{itemize}
   \item Tips, techniques, and training videos for communicating to various audiences.
   \item Fact sheets, brochures, and other handouts available for download.
   \item Talking points about the science of particle physics and its impact on society, with supporting examples and data.
   \item Links to existing websites and electronic materials that support communication, education, and outreach activities.
   \item Links to external databases or community-generated databases that track important statistics on workforce development, technology transfer, or economic impact.
\end{itemize}

Creating and maintaining such a website requires at least one FTE dedicated to nationwide particle physics communication, plus initial funding to create the website framework and continuing funding for website hosting and maintenance. The creation of particle physics-specific training videos and databases to track various types of statistics could require significantly more additional resources.

\textbf{Recommendation 3: Organize and identify logistical support for year-round campaigns in which particle physicists strategically advocate for scientific research with policy makers.}

Many attendees at the 2013 Community Summer Study meeting in Minneapolis expressed the desire for an ongoing effort that strategically leverages the widespread U.S. particle physics community to keep support high among policy makers. Existing efforts are mainly organized by users' groups and rely on scientist volunteers. Year-round strategic campaigns require additional logistical support dedicated to such a nationwide effort. A minimum of 0.5 FTE would be needed to:
\begin{itemize}
    \item Provide logistical support that enables the annual facility users groups' D.C. trip to be expanded to the whole U.S. particle physics community.
   \item Provide logistical support to organize a nationwide effort for physicists to visit their Congressional representatives in their home districts.
   \item Liaise with professional lobbying firms/organizations to organize nationwide letter-writing campaigns at key times during the federal budget cycle.
   \item Organize a nationwide effort to encourage every recent particle physics Ph.D. graduate from a U.S. university to write a personal letter to his/her Congressional delegation thanking them (if applicable) for the support that made their training possible, providing a brief overview of their research and its value to society, and describing where that student hopes to use their training.  
   \item Set up computers at key particle physics meetings and conferences at which scientists can send letters to their representatives.
\end{itemize}

\textbf{Recommendation 4: Provide communication training to the U.S. particle physics community.}

Communication training can greatly enhance scientists' ability to build support for their field, yet very few scientists have gone through such training. This subgroup recommends that the dedicated personnel identified in Recommendation 1: 
\begin{itemize}
    \item Identify existing training resources and make them available through the central communication resource identified in Recommendation 2.
   \item Investigate, together with the physics community, the most effective training format(s) and carry out pilot trainings.
   \item Organize at least one communication-related seminar or training at each major U.S. particle physics conference.
   \item Provide yearly assessments of this training and its impact on nationwide communication efforts.
\end{itemize}

\textbf{Recommendation 5: Work with APS to investigate the feasibility of a U.S. economic impact study for physics research that includes particle physics as a key component.}

Many policy makers and opinion leaders evaluate requests for funding based on an economic model of investment and return. The particle physics community recognizes this and seeks ways to compile such economic data for its research. Economic impact studies are relatively simple to conduct on behalf of an organizational entity such as a national laboratory or university, when the items studied are based on capital and operating costs such as federal funding, purchases of goods and services, payments of salaries to employees and expenditures by visitors. Studies carried out by CERN~\cite{CERNecstudy}  and the Canadian Light Source~\cite{CLSecstudy} have gone one step further, including some aspects of the economic and social impact of academic research, and the auxiliary effect of research activities and companies doing businesses with the laboratories on the region surrounding the laboratories. Studies that attempt to rigorously quantify the long-term impact of an entire field of fundamental science on a national economy are very rare, but there is no such study for particle physics. 

Due to the complexities and costs involved in such a study, this subgroup recommends that the first step for the particle physics community should be to partner with APS to investigate the feasibility of an economic impact study of physics research in the United States. This study should be balanced with respect to subfields of physics and identify connections of specified disciplines with particular industries. It should work toward an evaluation metric that includes weighted quantities in an attempt to better gauge the indirect economic impact of physics research. This study should also include the direct impact from the purchase of goods and services in particular regions as well as the impact of workforce development.

\textbf{Recommendation 6: Work with DPF and HEPAP to develop a sustainable process for collecting statistics on workforce development and technology transfer.}

Ongoing efforts exist to collect and analyze statistics on particle physics-related workforce development, in particular through the HEPAP Demography Committee and the HEPFolk database maintained at Lawrence Berkeley National Laboratory.  We recommend that resources be allocated to extend the work of this committee to set in place long-term, sustainable procedures for collecting and maintaining data on workforce training (including undergraduate training) and the jobs taken by Ph.D. physicists who do not pursue academic careers.

We further recommend that resources be allocated to create and initially populate a database of design and engineering patents that are associated with particle physics experiments. A simple search in August 2013 of the current U.S. patent collection database for the words ``particle physics" yields 497 hits~\cite{PTOdatabase}. Physicists will be able to use the database to generate reports for communication activities. 

\textbf{Recommendation 7: Generate letters and statements from third-party advocates in support of the impact of particle physics on society.}

Third-party advocates such as CEOs, notable scientists in other fields, and opinion leaders can be very powerful voices for particle physics research funding. Biologist Harold Varmus advocated for particle physics by pointing out that research instrumentation for biological science research was not developed by biologists, but by physicists~\cite{Varmus}.  We recommend that the U.S. particle physics community, together with the APS Division of Particles and Fields, identify and solicit support from CEOs of major companies in various U.S. industries in the form of letters of support for particle physics research that can be forwarded to policy makers. Similar letters should also be written and signed by the presidents of the major U.S. particle physics research universities. In conjunction with these communications, testimony before appropriate Congressional committees by CEOs, university presidents, and notable scientists from other fields should be arranged in support of particle physics research, facilities, and training. 
 
\section{Interconnections and common cause with scientists from other fields}

The field of particle physics has made fantastic advances over the past two decades. The discoveries of the top quark in 1995, tau neutrino in 2000 and a Standard Model Higgs-like boson in 2012 completed the Standard Model picture of particle physics. However, many other important developments such as the discovery of neutrino oscillations, discrepancies in some precision measurements of Standard Model processes, observation of matter-antimatter asymmetry, and
 the evidence for the existence of dark matter and dark energy all point to new physics beyond the Standard Model. The measured mass of the Higgs boson, while consistent with a SM particle, is also tantalizingly consistent with what could be expected from new physics such as supersymmetry. While much knowledge has been gained, there remain many intriguing questions to be addressed and a rich landscape of physics to be explored. 

As the field has marched on to probe deeper into the nature of fundamental particles and forces, the particle physics facilities and experiments have become larger, more complex and ever more expensive. The field of particle physics has become international and global. The United States, which dominated the field for many decades with many world-class high-energy particle accelerators in quick succession, is now facing a crisis with regards to maintaining such world leadership. The cancellation of the Superconducting Supercollider (SSC) project in the U.S. in 1993, the Large Hadron Collider at CERN superseding the Fermilab Tevatron as the world's Energy Frontier machine for particle physics, and the lack of a grand vision for U.S.-based particle physics facilities beyond the Tevatron have considerably weakened the U.S. position in the field, even while the field is growing stronger world-wide.

If particle physics is to remain viable and strong in the U.S., we need support from the public, policy makers, and from the broader science community. The goal of the subgroup on science community outreach was to review previous outreach activities our field has undertaken aimed at scientists in other fields and develop strategies for improving communication and outreach in the future.

Our broad goals for the subgroup through the Snowmass process were to explore ideas on how to build rapport with scientists from other physics subfields and other sciences, to get them to stand with us in areas of common cause and support a healthy particle physics program in the United States. In order to be able to communicate our science to the broader science community, it is also critical that the particle physicists themselves have broader appreciation and respect for all areas of particle physics and support the community plan to move the field forward. We therefore made efforts on both ``particle physics inreach" among the three established frontiers of particle physics (Energy, Intensity and Cosmic Frontiers) as well as ``particle physics outreach to scientists from other fields."

Here is a brief summary of our efforts and activities as a part of the Snowmass Community Summer Study over the past nine months.

\subsection{Past and current outreach activities}

We surveyed past and current outreach activities of the particle physics community targeted at colleagues in the broader science community. These include: (1) colloquia and  seminars at university departments and labs, (2) plenary sessions organized at APS and AAAS meetings, (3) publication of particle physics results in journals such as {\it Science} and {\it Nature}, (4) pedagogical review articles in similar journals, (5) articles in {\it Scientific American}, {\it Popular Science}, and similar magazines, (6) online articles and science blogs, (7) popular science books, and (8) reports written for a general audience commissioned by labs and agencies, for example, {\it Quantum Universe}~\cite{QuanU}, and {\it Discovering the Quantum Universe}~\cite{DiscQU}.

These activities have been useful in informing scientists in other fields about the latest work being done and results from particle physics, and these activities should be encouraged and continued. However, we conclude that more direct engagement with scientists in other fields is critical. 

\subsection{Fostering connections with other sciences}

Particle physics research has had a significant impact on other areas of science. Examples include applications developed for health and medicine, photon detection devices, neutron scattering, muon spin relaxation, large-scale computing and a variety of other fields.Ê Opportunities for improving connections in overlapping areas and for leading interdisciplinary efforts have to be embraced. To this end, the DOE Office of High Energy Physics has recently charged a ``Science Connections"Ê task force with highlighting the scientific areas where particle physics advances, informs and benefits from other DOE Office of Science programs. Such a report has not been attempted since the 1998 National Academy EPP Decadal Survey~\cite{EPP1998}.

\subsection{Survey results}

In the spring 2013 CE\&O survey two questions asked how particle physicists thought their colleagues in other fields perceived their field, and asked their opinions on the best way to broaden support for particle physics~\cite{PMOLsubgroup}. A large majority, 81\%, thought that particle physics is well or highly regarded by colleagues in other fields while only 17\% felt other scientists were indifferent or regarded particle physics negatively. A plurality, 46\%, felt that exploring interconnections with other disciplines combined with engaging in stronger outreach to scientists in other fields are two of the best ways to broaden support. 

\subsection{Particle physics inreach/outreach discussions}

Following discussions at the planning meeting held at Fermilab, October 10-12, 2012, we had several informal and formal discussions within the particle physics community and with non-particle physicists at meetings such as the AAAS meeting in Boston, February 2013, and the 2013 APS spring meetings in Baltimore and Denver. We held discussion sessions at the April APS meeting in Denver with invited speakers and a meeting at Fermilab regarding particle physics inreach and outreach. 

We organized three discussion sessions at the meeting in Minneapolis. The first was a panel session on ``Fostering Interconnections and Common Causes with Scientists in other Fields." This panel had a mix of particle physicists and condensed matter, astronomy, astrophysics and theoretical physics colleagues. At this session discussions led to ideas for future interactions between particle physics and scientists in other fields: (1) first and foremost there has to be mutual respect; (2) there are many opportunities for interconnections and interdisciplinary collaborations; (3) particle physics is often a technology incubator for other sciences; (4) particle physics has a lot to give and take provided that we exert the time and effort to break down barriers; (5) all fields in the United States are facing serious funding problems, and supporting each other will achieve better results for all sciences.

The second session was on ``Particle Physics Inreach and Communication Across the Three Frontiers of Particle Physics." The goal for this session was not only to discuss challenges and strategies for communication across frontiers within particle physics but to also facilitate a discussion on the need for all frontiers to weave a more unified story for particle physics.

Many within the community have voiced concern regarding the Venn diagram used for the visual representation of the three frontiers. While it is a tool for the funding agencies to promote and support a more modern view of the field, those concerned said that the resulting stovepiping threatens to effectively turn the field into three disjointed communities fighting for the same budget. The discussion on this topic elicited comments on the pros and cons of the frontier representation. In the end, there was a consensus that we need analogies such as these, maybe more of them rather than eliminating what we now have, and we need to be careful to use them in a positive way rather than to impose limits. The future success of our physics will require a program that is agile and maximizes the input from all the frontiers, even if in a temporally and spatially asynchronous way.

The third session was a community discussion on the HEPAP P5 prioritization process that is expected to produce a U.S. particle physics roadmap for the next one to two decades, with a report due before the summer of 2014. The particle physics community has been frequently told by representatives of the funding agencies that the community has to buy into the prioritization process and support the resulting plan. This community discussion provided an initial necessary opportunity for the particle physics community to express its views and provide input to the agency leaders and the HEPAP chair on the P5 panel and the process.

As a result of these many meetings, discussion sessions and deliberations, we developed the following strategies and implementation plans to help the particle physics community achieve the overarching CE\&O goals.

\subsection{Strategies for fostering connections and making common cause with other sciences}

\begin{enumerate}
 	\item Foster more dialog and understanding among subfields of physics and with other sciences.
	\item Identify areas of common cause and unite in support of them.
	\item Develop recognition for the need to prioritize critical research directions for the U.S. particle physics program; support the process and the resulting plan.
\end{enumerate}

\subsection{Implementation ideas}

\begin{enumerate}
 	\item Work with APS to create and foster new opportunities for dialog between leaders of physics subfields. 
	\item Hold combined open sessions at March/April APS meetings to disseminate the better understanding and the identified common causes.
	\item Continue the community discussion about the prioritization process so that the members of the community will support the resulting plan irrespective of how their favorite activities fared.
	\item With the help of APS, create and foster new opportunities for interaction with fields beyond physics.
\end{enumerate}

\section{Building public appreciation for particle physics}

Modern physics, like ancient physics, is driven by curiosity about the nature of our universe and of the world around us. By engaging the broader community in our research, we share this curiosity beyond the bounds of active physicists. The word ``engage" is central: we need to do more than inform the public: we need to engage them. It is critical that the public recognize the value of science in informed societal decision making.

The general public audience helps draw together the other CE\&O audiences. Many of the concepts developed for the general public apply to other audiences and successful communication to the general public reinforces communication to other audiences. When we communicate successfully with the general public, they will share their knowledge and enthusiasm with students, teachers, policy makers, the news media, and others.

Successful engagement with the public must be tailored to the audience. We have identified representative audiences meant to cover the range of situations where different engagement strategies and tactics are appropriate. Our messages have been categorized as follows: scientific goals, direct applications, and spinoffs and technology transfer. Each of the categories has a role to play in spanning the range of messages from human curiosity to economic impact. 

Within our community, many individuals and groups bring their enthusiasm for particle physics and its fundamental concepts to a wide range of audiences including the general public. They are aided by resources created and made available by their colleagues and communication professionals. These resources also impact other audiences including policy makers, other scientists, teachers, and students.

Among these resources are text materials (in books, brochures, websites), photos, images, animated and real-life videos, PowerPoint-type slides, artifacts and exhibits, recorded talks, hands-on activities, posters, smartphone apps, and art and science projects.

Communicating with the general public has great opportunities and some challenges. Much as we might wish to engage everyone in the general public, our reach is more generally limited to the 20-30\% of the public who are generally interested in science. These individuals come to us with very mixed backgrounds ranging from limited science literacy (with knowledge similar to that of a 14-15 year-old) to others who have read significantly about science and have a real foundation. Many come with broad misconceptions about the facts of physics, the nature of our research and even the basic principles of science. Others come with some strong skepticism that challenges our communication skills.

We usually have access to our audiences for very limited times, not the semester of lectures we have in universities. There are therefore strong limits to how much physics we can teach. We can be good at communicating the excitement and the impact of our research. We need to communicate the collaborative nature of modern research and the broad groups that come together to do experiments. The experiments frequently have collaborators from tens of countries who work together intensively and harmoniously with common goals. Some people are fascinated by the bottom-up nature of a 3000-person experiment, the opposite structure of corporations which also builds massive projects.

We need to portray the scientists who do these experiments as real and ordinary caring people. Communicating the impact of our research beyond particle physics is also essential. The technology transfer of modern physics has transformed the world and had an impact.

We have great stories to tell and should celebrate our successes and accomplishments. The discovery of the Higgs boson is a great story of three almost miraculous accomplishments, and U.S. physicists played vital roles in all three. The first was the development of the theory of the Higgs mechanism almost 50 years before its discovery. It is an inspirational example of the power of science to analyze complex problems. The second was the development of a 17-mile-long accelerator using revolutionary technologies. This accelerator then achieved a fantastic rate of collisions that was essential to make possible this discovery. And finally the enormous detectors achieved incredible levels of precision almost from the beginning, so that they could extract this discovery from the complex phenomena. The computing power required was also quite remarkable. In all aspects, U.S. physicists joined forces with physicists from dozens of countries in a wonderful example of global cooperation in a peaceful venture.

\subsection{Strategies for building appreciation for the value and excitement of particle physics among the general public}

The following strategies will achieve our overarching CE\&O goals:
\begin{enumerate}
 	\item Engage the public in a wide range of outreach activities. 
	\item Make the public aware of direct and indirect applications of research, both historical and potential. 
	\item Communicate the role and stories of U.S. physicists in particle physics, particularly in major discoveries and in the context of our international collaborations. 
\end{enumerate}

\subsection{Definitions of message categories}

\begin{itemize}
    \item Scientific goals: The quest for understanding the universe, including both the theoretical mysteries and the experimental challenges. 
   \item Applications: What applications might these discoveries have? Historical examples of unexpected applications of fundamental physics discoveries (e.g., quantum mechanics leading to the transistor and laser and special relativity leading to nuclear power).
   \item Spinoffs and technology transfer: Broader impacts of investing in science: (e.g., the web, grid/cloud based-computing, accelerators, and high-tech workforce).
\end{itemize}

\subsection{Audiences within the general public }

The message and format we use to engage the general public depends on the individual and on the context of the interaction. Below we establish representative audiences meant to cover the range of situations where different engagement strategies and implementations are appropriate. We are especially able to reach the first four categories.

\textbf{Popular science enthusiasts} are generally already convinced that science is exciting and valuable. They thirst for more information about the science, potential applications and the interaction with those actually doing the science.

\textbf{Everyday people} are engaged through natural social interaction ranging from a dinner party to an airplane flight. 

\textbf{Parents} may or may not be comfortable with or appreciative of science themselves, but they often support their child's interest in science and/or a scientific career path. 

\textbf{A geographic audience} consists of the populations near our universities and labs to which we have special access and opportunities.  

\textbf{Science skeptics} are often misinformed or have misconceptions about the fundamental scientific process. While many may be steadfast in their opinions, some have never been exposed to the counterarguments.

\textbf{Critics of public funding of science} are often indifferent to our scientific goals and seek specific examples of historical success and future prospects. 

\subsection{Survey results}

In the spring 2013 CE\&O survey 62\% of respondents self-identified as being engaged in CE\&O activities~\cite{PMOLsubgroup}. While we do not know what fraction of people in the field as a whole are actively involved in such activities, these respondents are considered to be disproportionately active in CE\&O activities.

The survey revealed that physicists think that the public is most interested in the potential application of discoveries followed by our scientific goals. The respondents reported that the general public is moderately interested in spin-off applications and least interested in a high-tech workforce, both of which belong to the third message category.

In terms of justification for the cost of particle physics research, the respondents indicated that the general public is most interested in potential direct applications of discoveries, followed by spin-off applications. 

The respondents are very comfortable communicating the overarching scientific goals of particle physics, but less comfortable in communicating the messages that are considered most interesting and which best justify the cost of particle physics research. Keeping in mind that the respondents are likely to be more involved in outreach than the field as a whole, this result informs which areas we as a field need to improve our communication.

\subsection{Existing communication and outreach activities to the general public}

The existing activities are extremely broad but with varying levels of support in terms of interest, participation and financial support. The single most common activity (perhaps because it requires limited financial support) is public talks. These occur in schools, clubs, university and lab events, scientific cafes, speaker bureaus and science theater-type events. 

Another common activity is participation in open houses and related events such as science festivals, lab and department tours, physics day shows, alumni weekends and workshops for the public. These require further effort but for limited periods.

Other activities take more time and require more resources, but also can have much more impact. These include contributions to external publications and shows. Examples include writing magazine articles and op-ed pieces in newspapers, participating as consultants to radio and television programs and movies and working with the news media. Individuals and institutions have collaborated with museums on exhibitions. In some cases these have included opportunities for the public to try simulated analysis of data. 

The production of printed products, web-based materials and multimedia products has been important for existing activities. Printed materials include books, brochures and posters. Websites can deliver news, background information, activities, images and video. The social media (blogs, Facebook, Twitter, YouTube) are an important aspect of a web presence. Labs, experiments and others have developed many videos, photo and image collections, smartphone apps, murals and even a 9,660-piece LEGO model. Physicists use these valuable resources in their individual efforts.

\subsection{Implementation notes for various audiences}

\textbf{Popular science enthusiasts} are generally already convinced that science is exciting and valuable. They thirst for more information about the science and potential applications and the interaction with those actually doing the science. Extending our current activities is the core of this effort. Additional efforts in developing plausible examples of future direct applications are in order.

\textbf{Everyday people} are engaged through the natural social interaction ranging from a dinner party to an airplane flight. Being prepared with a few talking points or one-minute physics explanations is very helpful, as are a handy brochure or similar item to share. The personal narrative and the broader scientific enterprise are effective approaches for engagement for this audience.

\textbf{Parents} may or may not be comfortable with or appreciative of science themselves, but they are often supporting their child's interest in science and/or evaluating a scientific career path.
\begin{itemize}
   \item Parents who are comfortable with science may wish to be more informed or have a point clarified so that they can directly educate their children. 
   \item Those who are less comfortable with science may wish to be informed so that they can better connect with a scientifically enthusiastic child (tactics: understanding the universe, famous/great scientists, fancy gadgets, etc). 
   \item Others may not want to be engaged directly but may collect materials for their children. 
\end{itemize}

\textbf{Science skeptics} are often misinformed or have misconceptions about the fundamental scientific process. While many may be steadfast in their opinions, some have never been exposed to the counter arguments. We need to meet these people where they are, in the movie theater, shopping mall or farmer's market. We need new ways to attract their attention, capture their interest, and engage them.

\textbf{Critics of public funding of science} are often often indifferent to our scientific goals and seek specific examples of historical success and future prospects. They often start by asking about direct applications, though one can pivot the conversation to spinoffs. Historical examples can be powerful in making the point of direct applications; however, additional efforts in developing plausible examples of future direct applications are in order.
 
\section{Increasing the presence of particle physics everywhere students learn}

An important resource for a healthy particle physics community is a talented and diverse workforce. Whether a student's career path leads them to graduate school and research or into other STEM-related fields, particle physics provides an excellent training ground. 

Working with learners of all ages, from school children to interested seniors, increases their excitement and interest in science. This engagement leads to better appreciation of the societal value of basic research, gaining advocates for our field. It also helps develop knowledge and critical thinking to help people make informed decisions in their daily lives on technological issues that affect society as a whole.

While particle physics generates a large degree of public interest, federal support for the field is showing a discouraging trend. We feel a public that is more scientifically literate will be more willing to support the costs of this field. A June 2013 survey of American attitudes toward science~\cite{Amspeaks} indicates that 74\% of respondents answered ``Strongly agree" (32\%) or ``Somewhat agree" (42\%) to the question: 

{\it Even if it brings no immediate benefits, basic scientific research that advances the frontiers of knowledge is necessary and should be supported by the federal government.}

The survey indicates broad support of the {\it idea} of federal support for basic research. The field must seek ways to turn this support of the idea of federal support into {\it increased} support. 

The three overarching goals for the CE\&O frontier articulate all of these elements. One way to realize these goals is an increased presence of particle physics concepts everywhere students learn: in school, at home, in the community and on the Web. Educators and students should learn directly from research scientists about their exciting work, using tools that are grounded in education research. There should be vertical integration of opportunities for students to be exposed to particle physics, in line with the vertical integration of the Next Generation Science Standards, from elementary school through high school~\cite{NGSS}. These opportunities should continue through the undergraduate level, with particle physics ``hooks" in introductory, education, and other non-major courses, and research opportunities not only for future physicists but for future and current teachers.

\subsection{Achievements since Snowmass 2001}

\begin{figure}[t]
\begin{center}
\includegraphics[width=.6\textwidth]{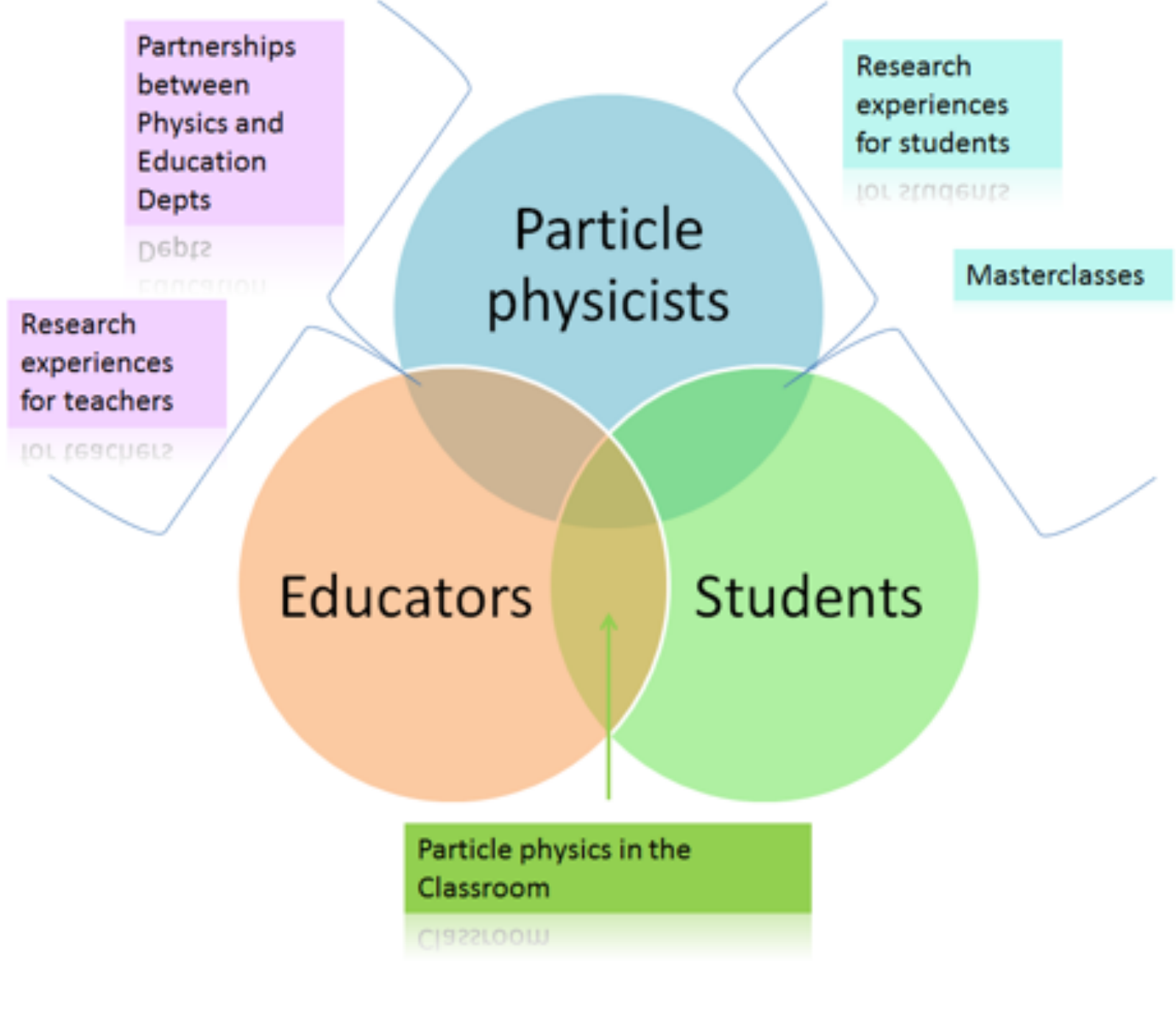}
\caption{A vision for inviting students and educators into the particle physics community.}
\label{default}
\end{center}
\end{figure}

The particle physics community has played an important role in workforce development with considerable progress in the past ten years. Opportunities for students to learn about physics by doing physics have blossomed as student enrollment at all levels has been increasing. Particle physics-related programs such as QuarkNet have been at the forefront. 

\textbf{K-12 education}

From 2001 to 2009, the number of students enrolled in physics in high school has grown from approximately 918,000 to approximately 1,350,000; while the percentage of girls has gone down slightly, the overall number of girls as well as boys has increased substantially~\cite{White2011}. AP Physics enrollment has also grown. At the same time, there is room for more growth, especially among women and minorities, which should start at the elementary level.

Particle physics offers tremendous opportunities for high school students. In QuarkNet, a long-term program embedded in the particle physics research community, students experience a research environment through summer research and cosmic-ray detector programs; they analyze particle physics data as physicists do in the International Masterclasses, e-Labs, and direct manipulation of 4-vectors in csv files.

The demand for physics high school teachers is enormous. Of all sciences taught at the secondary level, physics has the greatest shortage of qualified teachers: fewer than half of the 27,000 high school physics teachers in the U.S. have completed a major or minor in physics or physics education~\cite{White2010}. The particle physics community has made progress in creating meaningful professional development for teachers and career enhancement through research opportunities.

\textbf{Undergraduate education}

The number of physics bachelor's degrees awarded in the United States has been increasing steadily. According to an AIP report, the academic year 2010-11 produced new all-time highs for physics bachelor and Ph.D. degrees conferred in the United States~\cite{Nicholson2011}.

There has been great progress in research on physics education for undergraduates. Universities have access to curricula based on this research and best practices. These, when combined with the many available electronic aids, can have a positive impact on retention and student success. Faculty members are aided by the peer-reviewed journal {\it Physical Review Special Topics - Physics Education Research}. Dissemination of successful programs and important findings also occurs through the American Association of Physics Teachers (AAPT) and APS. 

Reform of undergraduate physics education can be successful with support from physics departments, both in terms of the leadership and the culture of the department. Particle physicists can and do play a role in such reforms through their own efforts and through their collaboration with teachers in education and outreach.

\textbf{Successful efforts}

The particle physics community in the United States and abroad has had success in increasing student interest and achievement in STEM fields, including particle physics. Examples include:
\begin{itemize}
   \item The successful University of Illinois undergraduate physics education reform initiative was recognized for its achievement with the APS 2013 Excellence in Physics Education Award~\cite{Smartphysics}.
   \item With 53 centers at universities and labs across the country, QuarkNet has put cosmic-ray detectors with online analysis tools and substantial data from LHC into the hands of high school students and teachers. Independent evaluation data from the long-term teacher professional development program show it has changed how many teachers see science and education. They also indicate it has helped students develop aspects of science literacy such as an understanding of how scientists make discoveries and talk about their work. Students develop this understanding through research internships, participating in masterclasses or conducting scientific investigations using e-Labs. 
   \item Netzwerk Teilchenwelt in Germany is one example where the QuarkNet model was adapted for students, teachers, and physicists in another country.
   \item The International Particle Physics Outreach Group (IPPOG) has successfully sponsored International Masterclasses for high school students since 2006. In 2013 there were 161 masterclasses in 37 countries, including 29 masterclasses from nine countries in the Fermilab-based portion of the program.
   \item In partnership with educators and physicists, the Fermilab Education Office has run successful education programs with particle physics and related physical science content for over 30 years. 
   \item The Contemporary Physics Education Project has brought together physicists and educators to create wall charts, posters, websites (e.g., ParticleAdventure.org and UniverseAdventure.org), and activities. The first efforts were in particle physics, and more recently in fusion and nuclear physics and cosmology. 
\end{itemize}

\newpage

\subsection{Issues and concerns }

\textbf{Funding}

The direct support of outreach efforts by particle physicists has been threatened in the 2014 budget by recent federal language to move support for K-12, undergraduate, and informal education programs from the agencies into three federal entities with very broad agendas. If this decision stands, it will threaten QuarkNet ½and other programs with DOE funding that connect scientists directly with students and teachers. 

\textbf{Technology}

The omnipresence of digital devices requires the particle physics community to explore the effective use of technology to deliver its messages. In addition to 1990's era webpages, the community must consider utilizing social media, mobile applications, data portals, video conferencing and other, not-yet realized tools as ways to carry out its outreach efforts. Particle physics was at the vanguard of the Web in the early 1990's; CE\&O should explore effective uses of this rapidly-evolving technology.
 
\textbf{Recognition for outreach efforts}

Young faculty members face many pressures on the path to tenure. These pressures often crowd out the important work of outreach. Physicists at other stages of their careers face different pressures. The particle physics community, and especially the university community, must change its culture such that  ---  both real and perceived --- are lowered for engagement in outreach efforts.

\textbf{Physics education}

The number of students enrolled in a high school physics course tripled from 1990-2005. If enrollment trends continue, the nation will need more people teaching physics. A recent survey of high school physics teachers asks what ``Resources [they] use to find answers about physics content." The top three resources were: college physics textbooks, the World Wide Web, and high school physics textbooks. The item ``Asking research scientist acquaintances" was down by an order of magnitude from these three. 

Improving physics instruction starts with helping teachers comprehend more about the field. This help can come from workshops and programs that build a teacher's instructional capacity and allow the teachers to learn from each other and from physicists. 

\textbf{Diversity}

In spite of numerous efforts, physics is still not a very diverse field. While women now comprise approximately 23\% of younger particle physicists~\cite{Anderson2013}), that number is still well below 50\%, and overall the number of women in the field is still 14\%. 

While there are many more girls taking high school physics, there are numerous reasons that they choose not to pursue physics majors in college. Working with girls in middle school and high school to introduce them to the range of careers available with a physics degree, and to female role models in physics, may aid in increasing the numbers entering the field. 

The number of students from underrepresented minorities in the field is still too low for meaningful statistics. Clearly, then, we need to attract more of these students into particle physics. In order to attract more students from underrepresented minorities, it is critical to work with the students in elementary school to strengthen their mathematics skills, and to interest students in pursuing mathematics and science courses and careers in STEM. We should then continue to provide opportunities for the students to engage science in an authentic, meaningful and exciting way.

\subsection{Strategies and implementation}

We encourage the direct engagement of physicists with students and educators at all levels from elementary through undergraduate. Particle physicists should invite educators and students into their unique community. Particle physicists should be involved in and support local, national and world-wide efforts that:

\begin{itemize}
\item{Offer long-term professional development and training opportunities for educators (including pre-service educators), aligned with current and appropriate standards and enabling educators to explore best-practice teaching methods. Make an effort to collaborate with local schools of education whenever possible.}
\item{Create learning opportunities for students of all ages, including classroom, out-of-school and online activities that allow students to explore particle physics to construct their own understanding and develop the skills and habits of mind necessary to perform research.}
\end{itemize}

The above strategies address all three goals of the CE\&O frontier through scientific literacy for all students and through inspiring and exciting the next generation of students entering particle physics and other STEM careers. These strategies might require commitments from physicists ranging from 15 minutes for an Ask-a-Scientist segment with a middle school student to a summer mentorship with a high school or college student to a multi-year relationship with a local teacher. All such commitments are equally valued and should be made as seamless as possible for the physicist. 

In order to fully realize these strategies, we call for the following implementation elements:

\textbf{Programs}

{\it Extend existing educator networks (e.g., QuarkNet, Modeling, PTRA) and create additional professional development opportunities to educators (including undergraduate students studying to be educators) who present science at all grade levels, especially those who teach physics. }QuarkNet is an excellent example of a program that fosters the connections between particle physicists and physics teachers, and, through the teachers, with students. It is successful in large part because it provides an infrastructure and framework that (1) fosters local learning communities of scientists and teachers doing science together, and (2) conveys an early understanding of the international nature of contemporary physics research to teachers and students. New programs will be piloted to expand professional development opportunities for teachers and reach more students.

{\it Create programs that use examples from particle physics and experimental data to allow educators and students to develop skills, concepts and scientific habits of mind.} Examples of such programs are University of Maryland's modern physics summer course for high school girls, and International Particle Physics Outreach Group (IPPOG) masterclasses where students analyze LHC data. Research skills workshops and research experiences for teachers, high school students and undergraduates are also important. 

{\it Align and collaborate with national efforts to improve diversity in STEM fields in order to broaden their reach in physics.} In order to have a healthy particle physics workforce in the future that more closely represents the diversity of the general public, students from groups traditionally underrepresented in physics need to be engaged and encouraged in elementary school. While direct content in particle physics may not be appropriate at this level, a general introduction to physics concepts can be. Several programs grounded in research on how students learn have gained national recognition in the past decade. These programs have not yet incorporated physics. We propose to partner with national and regional programs, for example SciGirls and GEAR-UP, to encourage the to include more physics content in their programs and materials. We will make those resources and programs known and available to interested members in the particle physics community. 

\textbf{Material resources}

{\it Produce learning materials that support the programs above, and materials for scientists that assist them in reaching a variety of audiences.} Groups such as the Contemporary Physics Education Project have produced particle physics posters and wall charts (including the latest on the Higgs discovery) and award-winning websites, such as the {\it Particle Adventure}~\cite{PartAd}, primarily through the effort of a few dedicated scientists and educators. Such efforts should be greatly extended to produce a variety of materials for a range of audiences, and these materials should be made available to scientists who wish to do outreach in their local communities. In addition, these materials should be made available to educators at both the K-12 and college levels to facilitate bringing best practices in physics education research to physics courses at all levels. 

\textbf{Training}

{\it Create workshops for scientists who want to engage educators and students.} A variety of workshops should be made available to scientists to who want to do outreach to schools or work with teachers, in order to keep them updated on education research and best practices, communication skills, etc. These could be coordinated in partnership with the APS and other professional organizations and made available to physicists in all fields.

\textbf{Resources}

Implementing these strategies requires a national infrastructure to provide an organizational structure and resources for scientists wishing to engage students and teachers as well as the general public. 
{\it  Establish such an organization within the particle physics community.} A national outreach office with a small, few-person national office could serve as a liaison office and resource center for particle physicists wishing to reach a variety of audiences. With the help of IT specialists and graphic artists, either in-house or outsourced, the staff would, for example, develop material that scientists could use for public lectures or school visits, offer training to scientists and partner with other organizations to develop diversity. Staff would facilitate collaborations among universities, national laboratories, and professional and international organizations to apply for funding for longer-term programs such as professional development and student enrichment programs.

\section{Conclusion}

As a result of the Community Summer Study process, the community has recognized that more physicists must engage in CE\&O activities, and that the quality and coordination of such activities must be improved, to increase public support for the field, develop the next generation of physicists, and ensure scientifically literate and engaged citizens.

A number of prominent voices reinforced the need for renewed commitment to CE\&O at the Minneapolis meeting:
\begin{itemize}
   \item We must educate our representatives in Congress, our fellow citizens, the business community and the scientific agencies. ---  D. Gross
   \item You are underselling yourselves\textellipsis you are technology incubators for other fields of science.   ---   R. Roser 
   \item The media missed the substantial impact of the U.S. on the Higgs discovery.  ---   J. Incandela
   \item You need to appeal to varied stakeholders to convince them that you do valuable science with a sensible plan. Illustrate the benefits of high-energy physics to society. ---   G. Blazey
\end{itemize}

The spring 2013 CE\&O survey of 641 members of the U.S. particle physics community identified the greatest barriers to participation in CE\&O activities as lack of time, little reward in terms of career advancement, and a lack of resources to communicate the broader societal impacts of particle physics research. Through collaboration with physicists, professionals in the areas of communication, government relations and education can expand the number of people engaged in education and public outreach by addressing these barriers.

Additional personnel and resources dedicated to nationwide coordination, training, resource development, evaluation, and support are required to achieve the goals and strategies outlined above. Such a team would enhance existing efforts and spearhead new initiatives such as: developing a comprehensive central CE\&O resource for physicists; developing sustainable methods to collect statistics on workforce development and technology transfer; producing materials to inform the public about direct and indirect applications of particle physics; and creating professional development opportunities for educators and new learning opportunities for students of all ages.




\end{document}